\documentclass{PoS}
\usepackage{cite}

\title{Measurement of deeply virtual Compton scattering (DVCS) cross sections with CLAS}

\ShortTitle{Measurement of deeply virtual Compton scattering (DVCS) cross sections with CLAS}

\author{\speaker{Hyon-Suk Jo}\thanks{for the CLAS Collaboration.}\\
        IPN Orsay, 91406 Orsay, France\\
        E-mail: \email{jo@ipno.in2p3.fr}}

\abstract{The generalized parton distributions (GPDs) provide a new description of the complex
internal structure of the nucleon in terms of its elementary constituents, the quarks and the
gluons. The GPDs describe the correlation between the transverse position and the longitudinal
momentum fraction of the partons in the nucleon, extending the information obtained from the
measurements of the form factors and the parton distribution functions. Deeply virtual Compton
scattering (DVCS), the electroproduction of a real photon from a single quark in the nucleon,
$eN \to eN\gamma$, is the most straightforward exclusive process that allows access to the GPDs.
A dedicated experiment to study DVCS with the CLAS detector at Jefferson Lab (JLab) has been
carried out using a 5.776~GeV polarized electron beam and an unpolarized hydrogen target,
allowing us to collect DVCS events in the widest kinematic range ever explored in the valence
region: $1<Q^2<4.6$~GeV$^2$, $0.1<x_B<0.58$, and $0.09<-t<3$~GeV$^2$. We present preliminary
results on the extraction of the DVCS/BH $ep \to ep\gamma$ cross sections.}

\FullConference{Sixth International Conference on Quarks and Nuclear Physics,\\
		April 16-20, 2012\\
		Ecole Polytechnique, Palaiseau, Paris}

\begin{document}

\section*{Introduction}

Generalized parton distributions (GPDs) take the description of the
complex internal structure of the nucleon to a new level by providing
access to, among other things, the correlations between the transverse
position and longitudinal momentum distributions of the partons in
the nucleon. They also give access to the orbital momentum contribution
of partons to the spin of the nucleon. 

GPDs can be accessed via deeply virtual Compton scattering (DVCS) and
exclusive meson electroproduction, processes where an electron interacts
with a parton from the nucleon by the exchange of a virtual photon, and
that parton radiates a real photon (in the case of DVCS) or hadronizes
into a meson (in the case of deeply virtual meson production). The
amplitude of the process can be factorized into a hard-scattering
part, exactly calculable in pQCD or QED, and a non-perturbative part,
representing the soft structure of the nucleon, parametrized by the GPDs.
At leading twist and leading order approximation, there are four
independent quark helicity conserving GPDs for the nucleon: $H$, $E$,
$\tilde{H}$ and $\tilde{E}$. These GPDs are functions depending on three
variables $x$, $\xi$ and $t$, among which only $\xi$ and $t$ are
experimentally accessible. The quantities $x+\xi$ and $x-\xi$ represent
respectively the longitudinal momentum fractions carried by the initial
and final parton. The variable $\xi$ is linked to the Bjorken variable
$x_{B}$ through the asymptotic formula: $\xi=\frac{x_{B}}{2-x_{B}}$.
The variable $t$ is the squared momentum transfer between the initial
and final nucleon. Since the variable $x$ is not experimentally
accessible, only Compton form factors (CFFs), ${\cal H}$, ${\cal E}$,
$\tilde{{\cal H}}$ and $\tilde{{\cal E}}$, whose real parts are weighted
integrals of GPDs over $x$ and whose imaginary parts are combinations of
GPDs at $x=\pm\xi$, can be extracted.

The reader is referred to Refs. \cite{gpd1, gpd2, gpd3, gpd4, gpd5,
gpd6, gpd7, gpd8, vgg1, vgg3, bmk} for detailed reviews of the
GPDs and the theoretical formalism.

\section*{Deeply Virtual Compton Scattering}

Among the exclusive reactions allowing access to GPDs, DVCS, which
corresponds to the electroproduction of a real photon off a nucleon
$eN \to eN\gamma$, is the key reaction since it offers the simplest,
most straightforward theoretical interpretation in terms of GPDs.

\begin{figure}[h]
  \centerline{\includegraphics[height=0.17\textheight]{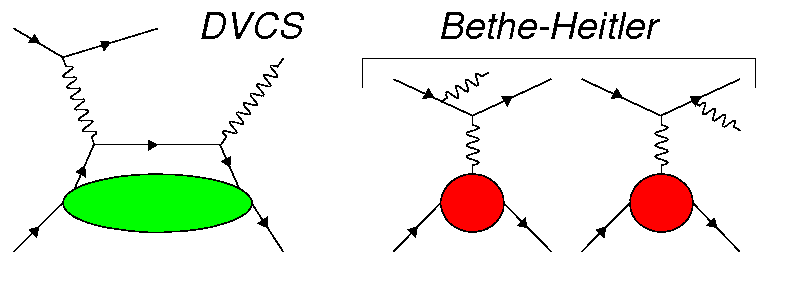}}
  \vspace{-0.2cm}
  \caption{One of the two handbag diagrams for DVCS (left) and diagrams
  for Bethe-Heitler (right). These two processes contribute to the
  amplitude of the $eN \to eN\gamma$ reaction.}
  \label{fig:diagrams}
\end{figure}

The DVCS amplitude interferes with the amplitude of the Bethe-Heitler
(BH) process which leads to the exact same final state. In the BH
process, the real photon is emitted by either the incoming or the
scattered electron while in the case of DVCS, it is emitted by the
target nucleon (see Figure~\ref{fig:diagrams}). Although these two
processes are experimentally indistinguishable, the BH process is
well-known and exactly calculable in QED. At current JLab energies
(6~GeV), the BH process is dominant in most of the phase space but the
DVCS process can be accessed via the interference term arising from the
two processes. With a polarized beam or/and a polarized target,
different types of asymmetries can be extracted: beam-spin asymmetry
($A_{LU}$), longitudinally polarized target-spin asymmetry ($A_{UL}$),
transversely polarized target-spin asymmetry ($A_{UT}$), and double-spin
asymmetries ($A_{LL}$, $A_{LT}$). Each type of asymmetry gives access
to a different combination of CFFs, being mostly sensitive to one or
two particular CFF(s). Single-spin asymmetries give access to the
imaginary part of the CFFs and double-spin asymmetries to their real
part. The DVCS/BH $eN \to eN\gamma$ unpolarized cross section is sensitive
to both the real part and the imaginary part of the CFFs. The polarized
cross section difference is linearly proportional to the imaginary part
of the CFFs.

\section*{The e1-DVCS experiment}

\begin{figure}[htb]
  \centerline{\includegraphics[height=0.3\textheight]{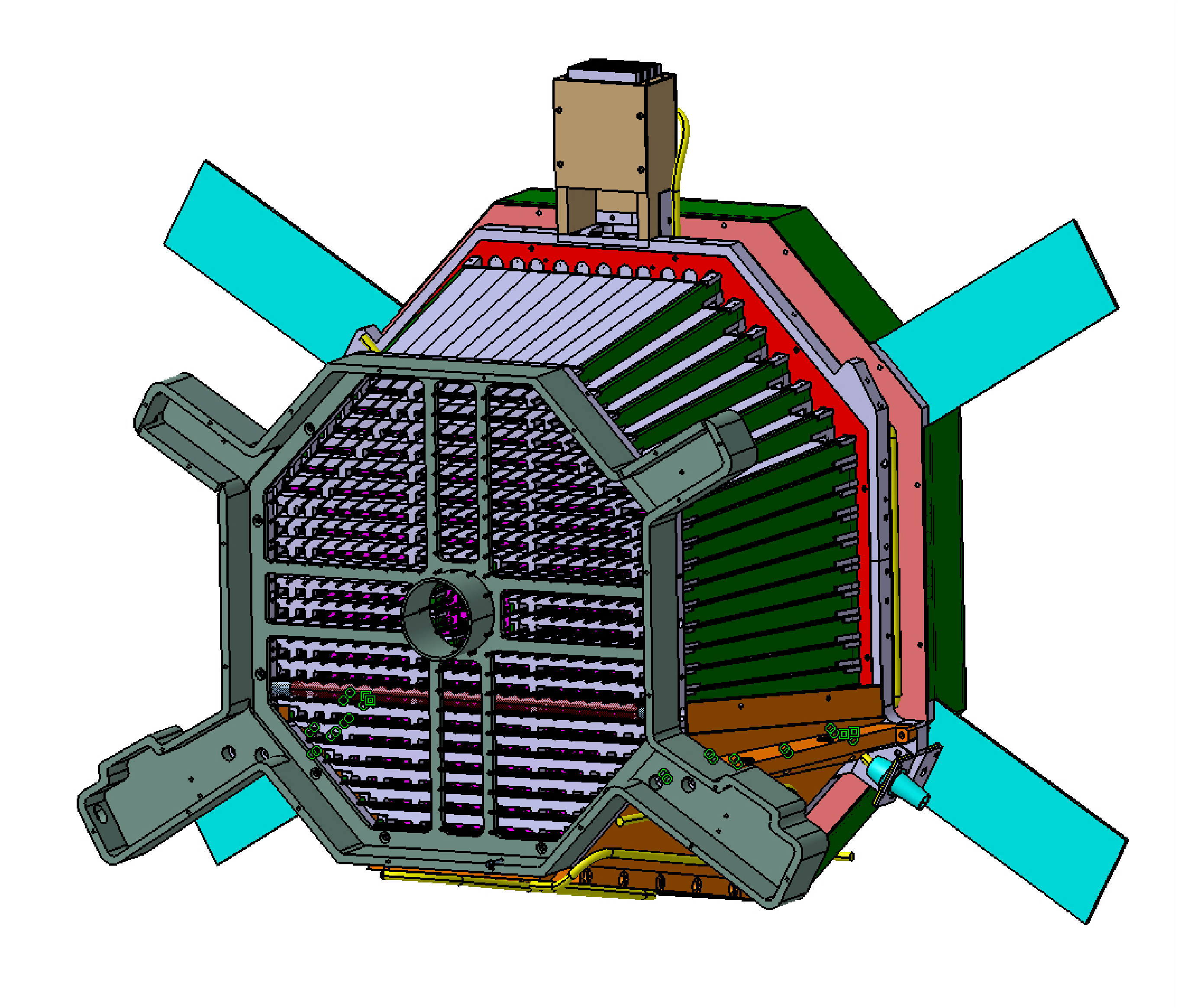}}
  \vspace{-0.4cm}
  \caption{The DVCS electromagnetic calorimeter, specially designed and
  built for the e1-DVCS experiment to detect DVCS/BH photons emitted at
  forward angles. It was later used in two other experiments dedicated to
  DVCS.}
  \label{fig:ic}
\end{figure}

The first DVCS results published by the CLAS collaboration were extracted
using data from non-dedicated experiments: beam-spin asymmetries in 2001
\cite{bsa1} and longitudinally polarized target-spin asymmetries in 2006
\cite{tsa}. In 2005, the first part of the e1-DVCS experiment was carried
out in Hall B at JLab using the CLAS spectrometer \cite{clas} and an
additional electromagnetic calorimeter, specially designed and built for
the experiment, and made of 424 lead-tungstate scintillating crystals read
out via avalanche photodiodes (see Figure~\ref{fig:ic}). This additional
calorimeter was located at forward angles, not covered by CLAS, where the
DVCS/BH photons are mostly emitted. This first CLAS experiment dedicated
to DVCS was fully exclusive and ran using a 5.766~GeV polarized electron
beam and a liquid-hydrogen target.

From the e1-DVCS data, CLAS published in 2008 the largest set of DVCS
beam-spin asymmetries ever extracted in the valence region \cite{bsa3}.
This report presents preliminary results on the extraction of DVCS/BH
$ep \to ep\gamma$ cross sections using the e1-DVCS data \cite{hsj}.

\section*{Extraction of DVCS/BH $ep \to ep\gamma$ cross sections}

\begin{figure}[h]
  \centerline{\includegraphics[width=0.68\textheight]{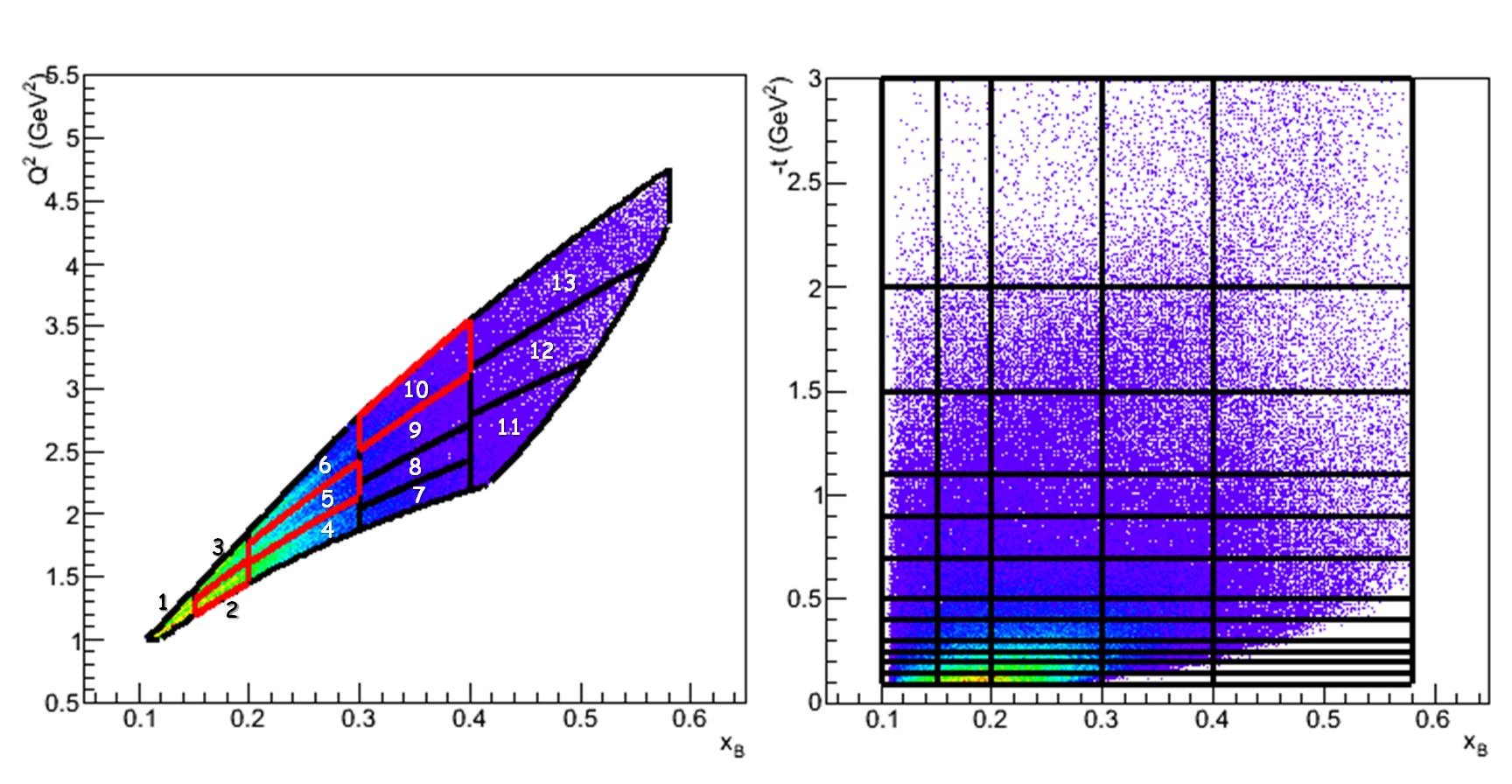}}
  \caption{Kinematic coverage of the e1-DVCS data. On the left: $Q^{2}$ as a
  function of $x_{B}$, showing the 13 bins in the ($Q^{2}, x_{B}$) plane chosen
  for the analysis. The bins highlighted in red (Bin 2, Bin 5 and Bin 10) are
  the ones for which results are shown in this report. On the right: $-t$ as a
  function of $x_{B}$, with the horizontal lines representing the 12 bins in $-t$.}
  \label{fig:coverage}
\end{figure}

The e1-DVCS data covers the range: $1<Q^2<4.6$~GeV$^2$, $0.1<x_B<0.58$, and
$0.09<-t<3$~GeV$^2$. We have analyzed the $ep \to ep\gamma$ reaction and are
in the process of extracting 4-fold differential cross sections
$\frac{d^{4}\sigma}{dQ^{2} dx_{B} dt d\Phi}$ in bins in ($Q^{2}$, $x_{B}$, $-t$,
$\Phi$). Figure~\ref{fig:coverage} shows the kinematic coverage of the e1-DVCS
data. On the left, is shown the phase space $Q^{2}$ as a function of $x_{B}$,
the lines representing the 13 bins in the ($Q^{2}, x_{B}$) plane chosen for
this analysis. Although we extracted the DVCS/BH $ep \to ep\gamma$ cross
sections for all bins, we show preliminary results only for three
representative bins (highlighted in red on the figure): Bin 2, at low $Q^{2}$
and $x_{B}$, Bin 5, at intermediate $Q^{2}$ and $x_{B}$, and Bin 10, at
relatively high $Q^{2}$ and $x_{B}$. On the right, is shown $-t$ as a function
of $x_{B}$, the horizontal lines representing the 12 bins in $-t$.

\vspace{-0.15cm}
\vspace{0.15cm}

\begin{figure}[hp]
  \centerline{\includegraphics[width=0.68\textheight]{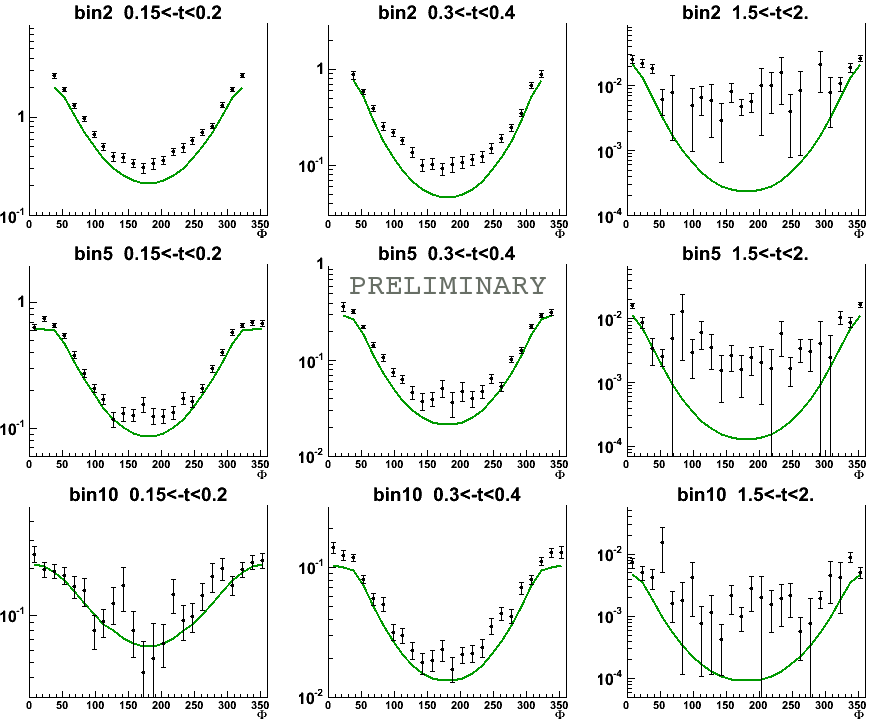}}
  \vspace{-0.2cm}
  \caption{Preliminary results for the $ep \to ep\gamma$ unpolarized cross sections
  (in nb/GeV$^{4}$) as a function of $\Phi$, for Bin 2, Bin 5 and Bin 10 and three
  bins in $-t$. The curves represent the BH calculation, integrated over each
  4-dimensional bin.}
  \label{fig:cs}
\end{figure}

\begin{figure}[hp]
  \centerline{\includegraphics[width=0.68\textheight]{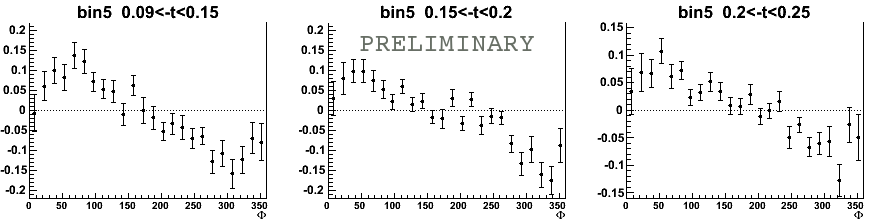}}
  \vspace{-0.2cm}
  \caption{Preliminary results for the $ep \to ep\gamma$ polarized cross section
  differences (in nb/GeV$^{4}$) as a function of $\Phi$, for Bin 5 and three bins
  in $-t$.}
  \label{fig:csdiff}
\end{figure}

We present preliminary results for the DVCS/BH $ep \to ep\gamma$ unpolarized
cross sections as a function of $\Phi$ (the angle between the lepton-scattering
plane and the hadronic plane) for Bin 2, Bin 5 and Bin 10 and three bins in $-t$
(see Figure \ref{fig:cs}). The results are shown with 24 bins in $\Phi$ although
the analysis was performed using 72 bins that were later merged. Only the
statistical errors are shown since systematic studies are still underway. The
curves represent the BH calculation integrated over each 4-dimensional bin. The
difference between the curves and the data can be attributed to the DVCS process
and the interference term, which are therefore a significant contribution. The
polarized cross section difference was extracted as well. Figure \ref{fig:csdiff}
shows preliminary results for Bin 5 and three bins in $-t$.

\section*{Summary}

DVCS/BH $ep \to ep\gamma$ unpolarized and polarized cross sections were extracted
in the widest kinematic range ever explored in the valence region. Preliminary
results for the unpolarized cross sections are compared to the corresponding BH
calculations. Preliminary results for the polarized cross section differences
are shown as well. These new results will allow the extraction of CFFs, using global
fitting codes, and will thus provide constraints on GPD models over a very large
kinematic range.

\section*{Acknowledgments}
Thanks to the organizing committee of the QNP2012 conference and to the
conveners of the parallel session for the opportunity to give this
presentation.

\end{document}